\newcommand{\reff}[1]{(\ref{#1})}

\newcommand{\of}[1]{\left( #1 \right)}
\newcommand{\norm}[1]{\ensuremath{\lVert #1 \rVert}}

\newcommand{\SNR}{\ensuremath{\mathsf{SNR}}}

\newcommand{\lmax}{\ensuremath{\lambda_{\text{max}}}}
\newcommand{\lmin}{\ensuremath{\lambda_{\text{min}}}}
\newcommand{\vmax}{\ensuremath{\mathbf{v}_{\text{max}}}}
\newcommand{\tr}{\ensuremath{\text{tr}}}
\newcommand{\diag}{\ensuremath{\text{diag}}}

\newcommand{\Rs}{\ensuremath{R_{\Sigma}}}
\newcommand{\RsT}{\ensuremath{R_{\Sigma,\text{TDMA}}}}
\newcommand{\Ro}{\ensuremath{R^{(1)}}}
\newcommand{\Rk}{\ensuremath{R^{(k)}}}

\newcommand{\Mk}{\ensuremath{M^{(k)}}}
\newcommand{\Mo}{\ensuremath{M^{(1)}}}
\newcommand{\Pk}{\ensuremath{P^{(k)}}}
\newcommand{\Pj}{\ensuremath{P^{(j)}}}
\newcommand{\Po}{\ensuremath{P^{(1)}}}

\newcommand{\A}{\ensuremath{\mathbf{A}}}
\newcommand{\B}{\ensuremath{\mathbf{B}}}

\newcommand{\F}{\ensuremath{\mathbf{F}}}
\newcommand{\Hrd}{\ensuremath{\mathbf{H}}}
\renewcommand{\P}{\ensuremath{\mathbf{\Pi}}}
\newcommand{\Fo}{\ensuremath{\mathbf{F}^{(1)}}}
\newcommand{\Fk}{\ensuremath{\mathbf{F}^{(k)}}}
\newcommand{\R}{\ensuremath{\mathbf{R}}}
\newcommand{\Rt}{\ensuremath{\mathbf{\widetilde{R}}}}
\newcommand{\I}{\ensuremath{\mathbf{I}}}
\newcommand{\U}{\ensuremath{\mathbf{U}}}
\newcommand{\Sg}{\ensuremath{\mathbf{\Sigma}}}
\newcommand{\So}{\ensuremath{\mathbf{\Sigma}^{(1)}}}
\newcommand{\La}{\ensuremath{\mathbf{\Lambda}}}
\newcommand{\V}{\ensuremath{\mathbf{V}}}
\newcommand{\Vo}{\ensuremath{\mathbf{V}^{(1)}}}
\newcommand{\Uo}{\ensuremath{\mathbf{U}^{(1)}}}
\newcommand{\Hkr}{\ensuremath{\mathbf{H}^{(k)}_r}}
\newcommand{\Hjr}{\ensuremath{\mathbf{H}^{(j)}_r}}
\newcommand{\Hro}{\ensuremath{\mathbf{H}^{(1)}_r}}

\newcommand{\Qk}{\ensuremath{\mathbf{Q}^{(k)}}}

\newcommand{\Qo}{\ensuremath{\mathbf{Q}^{(1)}}}

\newcommand{\bg}{\ensuremath{\boldsymbol{\tau}}}
\newcommand{\bgs}{\ensuremath{\boldsymbol{\tau}^{\ast}}}
\newcommand{\nus}{\ensuremath{\nu^{\ast}}}
\newcommand{\gk}{\ensuremath{\tau^{(k)}}}
\newcommand{\gkopt}{\ensuremath{\tau^{(k)}_{opt}}}
\newcommand{\gks}{\ensuremath{\tau^{(1)},\ldots,\tau^{(K)}}}
\newcommand{\ao}{\ensuremath{\alpha^{(1)}}}
\newcommand{\ak}{\ensuremath{\alpha^{(k)}}}
\newcommand{\aj}{\ensuremath{\alpha^{(j)}}}
\newcommand{\qo}{\ensuremath{q^{(1)}}}

\newcommand{\fo}{\ensuremath{f^{(1)}}}

\newcommand{\ve}{\ensuremath{\mathbf{v}}}
\newcommand{\x}{\ensuremath{\mathbf{x}}}
\newcommand{\xk}{\ensuremath{\mathbf{x}^{(k)}}}
\newcommand{\xr}{\ensuremath{\mathbf{x}_r}}
\newcommand{\yr}{\ensuremath{\mathbf{y}_r}}
\newcommand{\zr}{\ensuremath{\mathbf{z}_r}}
\newcommand{\hh}{\ensuremath{\mathbf{h}^H}}
\newcommand{\h}{\ensuremath{\mathbf{h}}}

\newcommand{\Qs}{\ensuremath{\mathbf{Q}^{(1)},\ldots,\mathbf{Q}^{(K)}}}

\newtheorem{theo}{Theorem}
\newtheorem{lemma}[theo]{Lemma}

\documentclass[conference,a4paper]{IEEEtran}

\ifCLASSINFOpdf
  \usepackage[pdftex]{graphicx}
  \graphicspath{{Figures/}}
  \DeclareGraphicsExtensions{.pdf,.jpeg,.png}
\else
  \usepackage[dvips]{graphicx}
  \graphicspath{{Figures/}}
  \DeclareGraphicsExtensions{.eps}
\fi

\usepackage{url}
\usepackage{cite}
\usepackage{amsmath, amsfonts}
\usepackage{pgf}
\usepackage{tikz}
\usepackage{pgfplots}
\usetikzlibrary{plotmarks}

\usepackage{subfigure}

\newlength{\myarraycolsep}
\newlength{\oldarraycolsep}
\begin{document}
\title{Superiority of TDMA in a Class of Gaussian Multiple-Access Channels with a MIMO-AF-Relay}

\author{\IEEEauthorblockN{Frederic Knabe, Omar Mohamed, and Carolin Huppert}
\IEEEauthorblockA{Institute of Communications Engineering\\
Ulm University, Albert-Einstein-Allee 43, 89081 Ulm, Germany\\
Email: \{frederic.knabe, omar.mohamed, carolin.huppert\}@uni-ulm.de}
}

\maketitle

\begin{abstract}
We consider a Gaussian multiple-access channel (MAC) with an amplify-and-forward (AF) relay, where all nodes except the receiver have multiple antennas and the direct links between transmitters and receivers are neglected. Thus, spatial processing can be applied both at the transmitters and at the relay, which is subject to optimization for increasing the data rates. In general, this optimization problem is non-convex and hard to solve. While in prior work on this problem, it is assumed that all transmitters access the channel jointly, we propose a solution where each transmitter accesses the channel exclusively, using a time-division multiple-access (TDMA) scheme. It is shown that this scheme provides higher achievable sum rates, which raises the question of the need for TDMA to achieve the general capacity region of MACs with AF relay.
\end{abstract}


% \fi
\IEEEpeerreviewmaketitle

\section{Introduction}
In today's wireless communication systems, the demand for higher data rates and wide-range coverage is steadily growing. To meet these requirements, a high density of base stations is necessary, which entails high costs for installation and maintenance. Another possibility to increase throughput and coverage is the use of relay nodes, which have much lower costs. Relay channels were considered in \cite{CG79} first, and have drawn more and more research attention in the last decades.

Depending on how the signals are processed at the relay, different types of relaying schemes are distinguished. The most common ones are amplify-and-forward (AF, also called non-regenerative relaying) and decode-and-forward (DF, also called regenerative relaying). While in AF, the relay simply amplifies the received signals subject to a power constraint, a complete decoding and re-encoding of the signal is necessary when using DF. As this yields higher costs and larger delays, we will restrict ourselves to AF relaying schemes in this paper.

In addition to relays, the deployment of nodes with multiple antennas helps to further increase the achievable data rates. The optimal combination of these two paradigms is an enormous challenge and has been considered in numerous publications, such as \cite{TH07, FHK06, JGH07} and the references therein. In \cite{TH07} an optimal amplifying matrix for a MIMO relay is found for the case where the transmit covariance matrices of the transmitting node is a scaled identity matrix. The general case with arbitrary transmit covariance matrices is considered in \cite{FHK06}, where the optimal structure of both relaying and transmit covariance matrix is found. However, this structure still contains parameters that are subject to optimization and the optimal solution of this problem remains unknown.

If multi-user systems are considered, finding the optimal transmit covariance and relaying matrices becomes even more complex. Achievable sum-rates for broadcast- \cite{CTHC08,YH10} and multiple-access channels (MAC) \cite{YH10} with relays have been derived, but the optimal rates have not been found yet.

In this work, we consider a $K$-user MAC with a full-duplex AF-relay, where all nodes except the receiver have multiple antennas. For this system, we first find the optimal solution for the approach adopted in \cite{YH10}. Subsequently, we will introduce a different transmission scheme based on time-division multiple-access (TDMA), that decomposes the $K$-user multiple-access relay channel (MARC) in $K$ orthogonal single-transmitter relay channels. Thus, the relay only receives the signal of one transmit station at a time, while, in the transmission scheme from \cite{YH10}, the relay receives all transmit signals jointly. Comparing the two schemes, we will show that the TDMA-based transmit scheme achieves sum-rates that are always larger or equal than those achieved by joint relaying of the signals in the considered channel.

This work is structured as follows: In Section \ref{sec:problem_formulation}, we introduce the channel model and describe the constraints that have to be fulfilled while optimizing the sum-rate. Subsequently, we review a previous solution of this problem in Section \ref{sec:joint_relaying} and optimize it for the model considered here. A different approach, based on TDMA, is proposed in section \ref{sec:TDMA}. After optimizing the parameters of the TDMA-based solution, its superiority is established. Subsequently, Section \ref{sec:simres} provides a quantitative comparison between the solutions by means of simulation results. Section \ref{sec:conclusion} concludes the paper.

\section{Channel model}\label{sec:problem_formulation}

\subsection{Notation}
We denote all column vectors in bold lower case and matrices in bold upper case letters. The trace and the Hermitian of a matrix $\A$ are identified by $\tr(\A)$ and $\mathbf{A}^H$, respectively. We use $\lVert \mathbf{x} \rVert$ to denote the Euclidean norm of a vector $\mathbf{x}$ and  $\I$ to describe the identity matrix. Furthermore, $\lmax(\A)$ and $\vmax(\A)$ indicate the largest eigenvalue of a matrix $\A$ and its corresponding eigenvector. A diagonal matrix with entries $a_1,\ldots,a_n$ is denoted by $\diag\of{a_1,\ldots,a_n}$.

\subsection{Channel Model}\label{subsec:channel_mod}

\begin{figure}
\centering{\includegraphics[width=\linewidth]{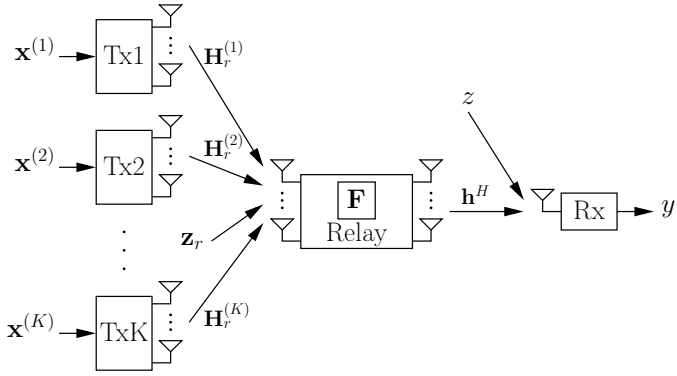}}
\caption{$K$-user multiple-access relay channel (MARC)}
\label{fig:MARC}
\vspace{-0.5cm}
\end{figure}
The MARC that we consider is depicted in Figure \ref{fig:MARC} and consists of $K$ transmitting nodes. User $k\in \{1,\ldots,K\}$ has $\Mk$ antennas, transmits the signal $\xk \in \mathbb{C}^{\Mk}$ to the relay, and has a channel matrix of $\Hkr \in \mathbb{C}^{M_r\times\Mk}$ for this transmission. Thus, the signal $\yr$ received at the relay can be written as
\begin{equation*}
\yr = \sum\limits_{k=1}^{K} \Hkr \xk + \zr,
\end{equation*}
where $\zr \sim \mathcal{CN}(0,\I)$ is the additive white Gaussian noise at the relay and $M_r$ denotes the number of antennas at the relay. The relay amplifies the signals by the matrix $\F$ and transmits the signal $\xr = \F\yr$ over the channel $\hh \in \mathbb{C}^{1\times M_r}$ to the receiver. For simplicity, it is assumed that the relay receives and transmits at the same time (full-duplex) and in the same frequency band\footnote{All results in this paper are also applicable to systems with relays that transmit at a different time (half-duplex) or in a different frequency band by normalizing the calculated rates with a prefactor accounting for the required extra time/bandwidth.} and that the direct paths between transmitters and receivers can be neglected. Hence, the received signal can be written as
\begin{align*}
y &= \hh \xr + z= \hh\F\yr + z\\
&= \hh\F\sum\limits_{k=1}^{K} \Hkr \xk + \hh\F\zr + z,
\end{align*}
where $z \sim \mathcal{CN}(0,1)$ denotes the Gaussian noise at the receiver. Both the transmitters and the relay are subject to average power constraints. We use $\Qk = E\of{{\xk\xk}^H}$ to denote the transmit covariance matrix of transmitter $k$. With this definition the transmit power constraints are given by
\begin{align}
\label{eq:PowerConstr}
\tr\of{\Qk} \leq \Pk \nonumber \;\; \forall k\in \{1,\ldots,K\} \\
E\!\of{\tr\of{\xr\xr^H}} \!=\! \tr\of{\!\F\!\of{\I+\sum\limits_{k=1}^{K} {\Hkr\Qk\Hkr}^H}\!\F^H\!} &\!\leq\! P_r. 
\end{align}
Finally, we assume that perfect channel state information is available at all nodes.

\section{Joint Relaying Scheme}\label{sec:joint_relaying}

In \cite{YH10} a transmission scheme was presented, where all transmitters send simultaneously and the relay amplifies and forwards the sum of all those signals. Therefore, we will refer to this scheme as ``joint relaying'' in the remainder of this paper. This scheme is briefly introduced in subsection \ref{subsec:previous}. In the following subsections, we first find the optimal relaying matrix $\F$ (\ref{subsec:opt_F}) and then derive the optimal covariance matrices $\Qs$ (\ref{subsec:opt_Q}).

\subsection{Previous Results}\label{subsec:previous}

The achievable sum-rate of the scheme from \cite{YH10} is given as
\begin{equation}
\label{eq:Rs}
\Rs = \log_2 \left(\, 1 + \frac{\hh\F \R \F^H\h}{\hh\F\F^H\h+1}\, \right),
\end{equation}
where $\R = \sum_{k=1}^{K} \Hkr\Qk{\Hkr}^H$, and it can be optimized with respect to $\F,\Qs$. Considering $\Hrd$ as a matrix of size $M_r \times N$, where $N$ is the number of receive antennas  \footnote[1]{$\Hrd$ is referred to as $\h$ in case of only one receive antenna ($N = 1$)}, its singular value decomposition (SVD) can be defined by $\Hrd = \U_h\Sg_h\V_h^H$ where $\Sg_h = \diag (\sigma_1,\ldots,\sigma_N)$ with descending diagonal order. Moreover, we define the eigenvalue decomposition of $\R$ as $\R=\U_r\La_r\U_r^H$ where $\La_r=\diag (\lambda_1,\ldots,\lambda_{M_r})$ with descending diagonal order. 

As stated in \cite{YH10} and shown in \cite{FHK06}, the optimum structure of $\F$ is given by
\begin{equation*}
\F = \U_h\Sg_f\U_r^H,
\end{equation*}
where $\Sg_f = \diag (f_1,\ldots,f_{M_r})^{1/2}$ and $f_i\geq 0\;\;\forall i$. 

\subsection{Optimal Relaying Matrix}\label{subsec:opt_F}
For any $\Qs$, we can reformulate \reff{eq:PowerConstr} and \reff{eq:Rs} with the expressions from above to obtain the following optimization problem (cf. \cite{YH10}):
\begin{align*}
\max_{f_1,\ldots,f_{M_r}} \Rs = \sum\limits_{i=1}^{min \left\{ N, M_r\right\}} \log_2 \left(1 + \frac{\sigma_i^2 \lambda_i f_i}{\sigma_i^2 f_i + 1} \right)\\
\text{s.t.}\;\; \sum\limits_{i=1}^{M_r} (\lambda_i+1)f_i \leq P_r \;\;\text{and}\;\; f_i \geq 0\;\;\forall i
\end{align*}
However, for the setup considered here, we have only $N=1$, i.e., $\sigma_1^2 = \norm{\h}^2$ and $\sigma_2=\ldots=\sigma_n=0$. As only $f_1$ contributes to the sum-rate, the problem is obviously  solved by $f_1 = P_r(\lambda_1 + 1)^{-1}$, $f_2=\ldots=f_{Mr}=0$ for this case. Hence, the sum-rate can be expressed as
\begin{align}
\Rs &= \log_2 \left( 1 + \frac{\sigma_1^2\lambda_1 P_r}{\sigma_1^2 P_r + \lambda_1 + 1}\right)\nonumber\\
%&= \log_2 \left( 1 + \frac{\norm{\h}^2\lmax(\R)P_r}{\norm{\h}^2P_r + \lmax(\R) + 1} \right)\\
&= \log_2 \left(1+\norm{\h}^2P_r\right) - \log_2 \left(1 + \frac{\norm{\h}^2P_r}{1+\lmax(\R)}\right)\label{eq:Rs_nonTDMA},
\end{align}
and can now be optimized with respect to $\Qs$.

\subsection{Optimal Covariance Matrices}\label{subsec:opt_Q}

As it can be seen from the last line of \reff{eq:Rs_nonTDMA}, the optimal sum-rate is achieved by maximizing the largest eigenvalue of $\R$. As the maximum eigenvalue operation is a convex function, the maximization is not trivial. To provide a solution, we will introduce two Lemmas:\pagebreak
\begin{lemma}\label{lem:ev_pd}
Let $\A,\B \in \mathbb{C}^{n\times n}$ be positive semidefinite matrices. Then, if $\A-\B$ is positive semidefinite ($\A \succeq \B$), we have $\lmax\of{\A} \geq \lmax\of{\B}$
\end{lemma}
\begin{IEEEproof}
The proof is given in Appendix A.
\end{IEEEproof}
\begin{lemma}\label{lem:pee_pi}
Let $\P \in \mathbb{C}^{n\times n}$ be a positive semidefinite matrix with $\tr(\P) \leq P$ and $\A \in \mathbb{C}^{m\times n}$ . Then, it holds that
\begin{equation*}
\A\A^HP \succeq \A\P\A^H
\end{equation*}
\end{lemma}
\begin{IEEEproof}
The proof is given in Appendix B.
\end{IEEEproof}
With these two lemmas, we can now formulate:
\begin{theo}
The optimal value of the non-convex optimization problem
\begin{align*}
\max_{\Qk}\;\;&\lmax\of{\R}\\
\text{s.t.}\;\;\; &\tr\of{\Qk} \leq \Pk \;\; \forall k
\end{align*}
is $\lmax\of{\sum_{k=1}^K \Hkr{\Hkr}^H \Pk}$.
\end{theo}
\begin{IEEEproof}
Define $\Rt = \sum_{k=1}^K \Hkr{\Hkr}^H \Pk$. Using Lemmas \ref{lem:ev_pd} and \ref{lem:pee_pi}, it is directly seen that $\lmax(\R) \leq \lmax\left(\Rt\right)$. For showing the achievability of this solution, let
\begin{equation}\label{eq:Qk_opt}
\Qk = \frac{{\Hkr}^H\ve\ve^H\Hkr}{\ve^H\Hkr{\Hkr}^H\ve} \Pk
\end{equation}
with $\ve = \vmax(\Rt)$. Apparently, $\tr\of{\Qk} \leq \Pk$ is satisfied and we can write
\begin{align}
\lmax(\R) &= \vmax(\R)^H \of{\sum\limits_{k=1}^{K} \Hkr\Qk{\Hkr}^H}\vmax(\R)\\
&\geq \ve^H \of{\sum\limits_{k=1}^{K} \Hkr\Qk{\Hkr}^H}\ve\label{eq:vRV}\\
&= \sum\limits_{k=1}^{K} \ve^H \Hkr {\Hkr}^H \Pk \ve  = \lmax\left(\Rt\right)\label{eq:lmaxRt},
\end{align}
where \reff{eq:vRV} follows from the Rayleigh quotient and \reff{eq:lmaxRt}
 is obtained by using \reff{eq:Qk_opt} and writing $\ve$ and $\ve^H$ inside the sum. Thus, choosing $\Qk$ as in \reff{eq:Qk_opt} leads to the optimal value $\lmax\left(\Rt\right)$.
\end{IEEEproof}

With this optimal choice of the covariance matrices the achievable sum-rate for joint relaying can be expressed as
\vspace{-0.1cm}
\begin{equation}\label{eq:Rs_jr_opt}
\Rs = \log_2 \left(1+\norm{\h}^2P_r\right) - \log_2 \left(1 + \frac{\norm{\h}^2P_r}{1+\lmax\left(\Rt\right)}\right).
\end{equation}

\section{TDMA-Based Relaying}\label{sec:TDMA}

In this section, we will introduce a relaying scheme based on TDMA. This scheme includes a division of the transmission in $K$ time slots, where  user $k$ occupies the $k$-th time slot exclusively. Also the relay incorporates this slot structure, i.e., the relaying matrix $\Fk$ in time slot $k$ can be adapted to the channel of user $k$ only. Thus, the TDMA slot structure decomposes the channel in $K$ independent single user relay channels. Therefore, we will first derive the optimal structure of the transmit covariance- and relaying matrix for the single user relay channel in subsection \ref{subsec:single_relay}. In the following subsection \ref{subsec:transmission}, we will transfer this scheme to the MARC with TDMA and optimize the achievable sum-rate. Finally, in subsection \ref{subsec:evaluation} we will show that this sum-rate is always larger or equal than the one in \reff{eq:Rs_jr_opt}.

\subsection{Single-User Relaying}\label{subsec:single_relay}

In order to describe a single-user relay channel with a consistent notation, we assume the same channel model as introduced in subsection \ref{subsec:channel_mod} and assume there is only $K=1$ transmitting user. For this scenario, it was shown in \cite[Theorem 1]{FHK06}, that the optimal structure of the source covariance matrix $\Qo$ and the relay matrix $\Fo$ are given by
\begin{align*}
\Qo &= \Vo \So_q {\Vo}^H\\
\Fo &= \U_h \So_f {\Uo}^H,
\end{align*}
where $\Hro = \Uo \So {\Vo}^H$ is the SVD of $\Hro$, $\So = \diag(\ao_1,\ldots,\ao_{\min\{\Mo,M_r\}})^{1/2}$, 
$\So_q = \diag(\qo_1,\ldots,\qo_{\Mo})$, and $\Sg_f = \diag(\fo_1,\ldots,\fo_{M_r})^{1/2}$.

With these expressions, the achievable (sum) rate $\Ro$ of the only user can be derived by evaluating \reff{eq:Rs}, which, together with the power constraints \reff{eq:PowerConstr} allows the formulation of an optimization problem (cf. \cite{FHK06}):
\begin{align*}
\max_{\begin{matrix}
\fo_1,\ldots,\fo_{M_r}\\
\qo_1,\ldots,\qo_{\Mo}
\end{matrix}} \Ro = \sum\limits_{i=1}^{\widetilde{N}} &\log_2 \left(1 + \frac{\sigma_i^2 \ao_i \qo_i \fo_i}{\sigma_i^2 \fo_i + 1} \right)\\
\text{s.t.}\;\; \sum\limits_{i=1}^{M_r} (\ao_i \qo_i+1)\fo_i &\leq P_r,\;\;\;\;\, \fo_i \geq 0\;\;\forall i\\
\sum\limits_{i=1}^{\Mo}\qo_i &\leq \Po,\;\;\qo_i \geq 0 \;\;\forall i,
\end{align*}
where $\widetilde{N} = \min \left\{ N, M_r,\Mo\right\}$.

In the case of $N=1$ receive antenna considered here, the problem is even simpler to solve. As in the joint relaying scheme, we have $\sigma_1^2 = \norm{\h}^2$ and $\sigma_2=\ldots=\sigma_n=0$. Thus, the optimal choice of $\Fo$ is again obtained by choosing $\fo_1 = P_r(1+\ao_1 \qo_1)^{-1}$, $\fo_2=\ldots=\fo_{M_r}=0$. Also for the transmit covariance matrix, only $\qo_1$ has a contribution to the rate, which makes $\qo_1 = \Po$, $\qo_2=\ldots=\qo_{\Mo}=0$ the optimal solution.
Using this solution, the achievable rate $\Ro$ can be written as
\begin{align*}
\Ro &= \log_2 \left( 1 + \frac{\norm{\h}^2\ao_1\Po P_r}{\norm{\h}^2P_r + \ao_1\Po + 1} \right)\nonumber\\
&= \log_2 \left(1+\norm{\h}^2P_r\right) - \log_2 \left(1 + \frac{\norm{\h}^2P_r}{1+\ao_1\Po}\right).\label{eq:R_single}
\end{align*}

\subsection{TDMA-based Transmission Scheme}\label{subsec:transmission}

In order to decompose the $K$-user MARC in $K$ single-user relay channels, we use a TDMA scheme, such that user $k$ transmits only in a time slot of duration $\gk$ with $\sum_{k=1}^{K}\gk = 1$. In each time slot, the optimal choice of the relay matrix $\Fk$ and the transmit covariance matrix $\Qk$ can be obtained as in subsection \ref{subsec:single_relay}. The only difference is the transmit power constraint: As user $k$ only transmits in $\gk$ fraction of the time, it can use a transmit power of $\Pk/\gk$ and still fulfills the average transmit power constraint \reff{eq:PowerConstr}. Thus, the rate of user $k$ is given by
\begin{equation*}
\Rk = \gk \!\left[\log_2 \left(1\!+\!\norm{\h}^2P_r\right) - \log_2\! \left(1\!+\! \frac{\norm{\h}^2P_r}{1+\ak_1\frac{\Pk}{\gk}}\right)\!\right]\!,
\end{equation*}
and the sum-rate can be calculated as $\RsT = \sum_{k=1}^K \Rk$. This sum-rate can be optimized by the choice of $\gks$, where the optimum is given as follows:
\begin{theo}\label{theo:optimal_gk}
For the considered TDMA transmission scheme in the $K$-user MARC, the optimal durations of the time slots $\gks$ are given by
\begin{equation}\label{eq:gkopt}
\gkopt = \frac{\ak_1\Pk}{\sum\limits_{j=1}^{K} \aj_1 \Pj}.
\end{equation}
\end{theo}
\begin{IEEEproof}
Consider the optimization problem
\begin{align*}
\max_{\gks} &\RsT(\bg)\\
\text{s.t.}\;\;&h(\bg) = 1 - \sum_{k=1}^K \gk = 0,
\end{align*}
where we introduced the vector $\bg = \left[\gks \right]$ and wrote $\RsT$ as a function of $\bg$ to be more consistent with the notation used in optimization theory. First, it is easy to show that the above problem is convex and thus the Karush-Kuhn-Tucker (KKT) conditions provide necessary and sufficient conditions for optimality. For the underlying problem, the KKT conditions of a solution $\bgs$ to be optimal can be formulated as $h(\bgs) = 0$ and $\nabla\RsT(\bgs) + \nus \nabla h(\bgs) = 0$, where $\nus \in \mathbb{R}$ can be chosen arbitrarily. If we choose $\bgs$ as suggested in Theorem \ref{theo:optimal_gk}, the first condition is obviously fulfilled. For the second condition we can calculate the derivatives of $\RsT$ and $h$ at the point $\bg = \bgs$ as
\begin{align*}
\left.\frac{\partial\RsT}{\partial\gk}\right|_{\bg=\bgs} &= \log_2 \left(1+\frac{\norm{\h}^2P_r}{S(1+\norm{\h}^2P_r)+1} \right)\\[0.2cm]
&\hspace{-1.5cm}-\frac{\log_2(e)S\norm{\h}^2P_r(1+\norm{\h}^2\Pr)}{(S(1\!+\!\norm{\h}^2P_r)\!+\!1)^2\left(1+\frac{\norm{\h}^2P_r}{(S(1+\norm{\h}^2P_r)+1)} \right)}\\[0.3cm]
\left.\frac{\partial h}{\partial\gk}\right|_{\bg=\bgs} &= -1,
\end{align*}
where $S=\left( \sum\limits_{k=1}^{K} \ak_1 \Pk \right)^{-1}$. As the derivatives with respect to $\gk$ are the same in all components $k=1,\ldots,K$, we can select $\nus = \left.\frac{\partial\RsT}{\partial\gk}\right|_{\bg=\bgs}$ and also the second KKT condition is fulfilled.
\end{IEEEproof}
Using this optimal time slot durations, the achievable sum-rate of the TDMA-based transmission scheme is given by
\begin{equation}\label{eq:RsTDMA}
\RsT \!=\! \log_2 \left(1\!+\!\norm{\h}^2P_r\right) - \log_2\! \left(1\!+\! \frac{\norm{\h}^2P_r}{1+\sum\limits_{j=1}^{K} \aj_1 \Pj}\right)\!.
\end{equation}

\subsection{Comparison with Joint Relaying}\label{subsec:evaluation}
The comparison between the proposed TDMA-based and the previously considered joint relaying scheme is given by the following theorem:

\begin{theo}\label{theo:TDMA_better}
In the considered $K$-user MARC, the TDMA-based transmission scheme with optimal time slot durations $\gk$ achieves at least the same sum-rate as the joint relaying scheme, i.e.,
\begin{equation*}
\RsT \geq \Rs.
\end{equation*}
\end{theo}
\begin{IEEEproof}
Comparing the rate expressions of $\RsT$ and $\Rs$ from \reff{eq:RsTDMA} and \reff{eq:Rs_jr_opt}, respectively, it can be seen that it is sufficient to show that $\sum_{j=1}^K \aj_1\Pj \geq \lmax\left(\Rt\right)$. This is easily obtained by writing
\begin{align}
\sum_{j=1}^K \aj_1\Pj &= \sum_{j=1}^K \lmax\of{\Hjr{\Hjr}^H\Pj}\\
&\geq \lmax\left(\Rt\right)\label{eq:lmax_R},
\end{align}
where \reff{eq:lmax_R} follows from the convexity of the maximum eigenvalue operation.
\end{IEEEproof}
Note that (cf. \cite{KT01}) in \reff{eq:lmax_R} equality holds only if all eigenvalues $\lmax\of{\Hjr{\Hjr}^H\Pj}$ have the same eigenvector. However, if the matrices $\Hjr$ are statistically independent, which is a valid assumption in practical systems, this happens with probability zero. The consequence of this is that the TDMA-based approach is strictly better.

A plausible explanation for this is the fact, that in joint relaying, the relay always has to come to a compromise, such that not only one but all incoming signals are amplified as good as possible. In the TDMA system, the relay has only one incoming signal, which can be optimally amplified by optimizing the relay matrix just for this signal.

\section{Simulation Results}\label{sec:simres}
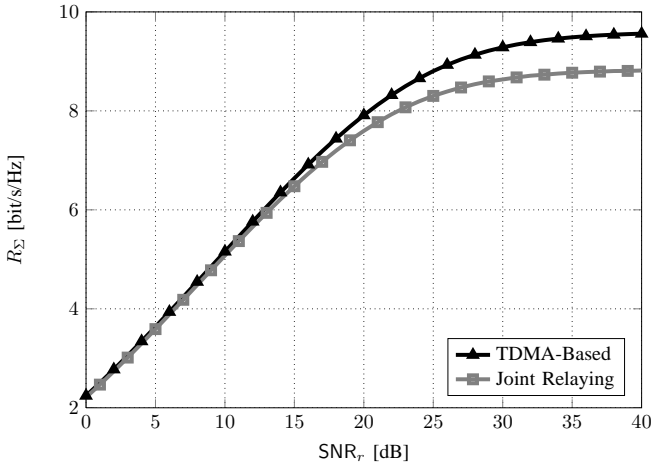
\begin{figure}
\centering{% This file was created by matlab2tikz v0.1.3.
% Copyright (c) 2008--2011, Nico Schlömer <nico.schloemer@gmail.com>
% All rights reserved.
% 
% The latest updates can be retrieved from
%   http://www.mathworks.com/matlabcentral/fileexchange/22022-matlab2tikz
% where you can also make suggestions and rate matlab2tikz.
% 
\begin{tikzpicture}[scale=0.75]
%\pgfplotsset{major grid style={dotted}}

\begin{axis}[%
scale only axis,
width=1.1\linewidth,
height=7cm,
xmin=0, xmax=40,
ymin=2, ymax=10,
xlabel={$\SNR_r$ [dB]},
ylabel={$\Rs$ [bit/s/Hz]},
xmajorgrids,
ymajorgrids,
zmajorgrids,
major grid style={color=black,dotted},
legend entries={TDMA-Based,Joint Relaying},
legend style={nodes=right},
legend pos=south east
]

\addplot [
color=black,
solid,
mark=triangle,
mark options={color=black},
mark phase=1,
mark repeat=2,
line width=2.0pt
]
coordinates{
 (-5,1.15202)(-4,1.33787)(-3,1.54092)(-2,1.76031)(-1,1.99489)(0,2.24328)(1,2.50398)(2,2.77543)(3,3.05604)(4,3.34428)(5,3.63868)(6,3.93784)(7,4.24045)(8,4.54526)(9,4.85106)(10,5.15667)(11,5.46089)(12,5.76249)(13,6.06018)(14,6.35259)(15,6.63828)(16,6.91571)(17,7.1833)(18,7.43945)(19,7.68258)(20,7.91124)(21,8.12416)(22,8.32033)(23,8.4991)(24,8.66015)(25,8.8036)(26,8.92993)(27,9.03996)(28,9.13477)(29,9.21566)(30,9.28403)(31,9.34133)(32,9.38898)(33,9.42832)(34,9.46062)(35,9.48698)(36,9.5084)(37,9.52574)(38,9.53973)(39,9.55098)(40,9.56001)(41,9.56725)(42,9.57303)(43,9.57765)(44,9.58134)(45,9.58427)(46,9.58661)(47,9.58848)(48,9.58996)(49,9.59114)(50,9.59208) 
};

\addplot [
color=gray,
solid,
mark=square,
mark options={color=gray},
mark phase=2,
mark repeat=2,
line width=2.0pt
]
coordinates{
 (-5,1.13156)(-4,1.31494)(-3,1.51547)(-2,1.73233)(-1,1.96436)(0,2.21021)(1,2.46825)(2,2.73695)(3,3.01453)(4,3.29941)(5,3.59)(6,3.88474)(7,4.18215)(8,4.48059)(9,4.77864)(10,5.07512)(11,5.36852)(12,5.65685)(13,5.93893)(14,6.21301)(15,6.47727)(16,6.72997)(17,6.96996)(18,7.19508)(19,7.40492)(20,7.5954)(21,7.77185)(22,7.93099)(23,8.07181)(24,8.19486)(25,8.30231)(26,8.39131)(27,8.47261)(28,8.53918)(29,8.59554)(30,8.63484)(31,8.67627)(32,8.71039)(33,8.73125)(34,8.75405)(35,8.7708)(36,8.78394)(37,8.79323)(38,8.80512)(39,8.80727)(40,8.81549)(41,8.81767)(42,8.82281)(43,8.82907)(44,8.82863)(45,8.83267)(46,8.83048)(47,8.83301)(48,8.83235)(49,8.83356)(50,8.83425) 
};
\end{axis}
\end{tikzpicture}}
\vspace{-0.5cm}
\caption{Comparison of TDMA-based and joint relaying scheme for $K=8$, $\Mo = \ldots = M^{(8)} = M_r = 4$, $\Po = \ldots = P^{(8)} = 10$}
\label{fig:TDMAvsSP}
\vspace{-0.3cm}
\end{figure}
In order to confirm the result of Theorem \ref{theo:TDMA_better}, we will evaluate the achievable rates of the TDMA-based and joint relaying scheme by some simulations in this section. As stated in the preceding section, the TDMA-based scheme is strictly better if the channel matrices $\Hjr$ are statistically independent. However, the question for the quantitative gain of the TDMA-based scheme has to be investigated.

For this purpose, we assume channels with independent Rayleigh-fading, i.e., all entries of the channel matrices\linebreak are $\sim \mathcal{CN}(0,1)$ and independent from each other. All results presented in this section are obtained by averaging over 1000 channel realizations.

For a system with $K=8$ users, the achievable sum-rates are plotted in Figure \ref{fig:TDMAvsSP}. We assumed that all transmitters and the relay have $\Mk = M_r = 4$ antennas, while, as a prerequisite of this scenario, the receiver is assumed to have a single antenna. The transmit power of the users is assumed to be fixed at $\Pk = 10$ $\forall k$, while the power of the relay $P_r$ and thus the corresponding signal-to-noise ratio $\SNR_r = P_r$ is varied. It can be observed that, especially for high values of $\SNR_r$, the performance of the TDMA-based scheme is up to $10$\% better than that of the joint relaying scheme. From further simulations, which are not visualized in this work due to the page constraint, it could be seen that this gain grows with the number of antennas at the transmitters $\Mk$, the number of antennas at the relay $M_r$, and with the number of users $K$.

\section{Conclusion}\label{sec:conclusion}
We have considered a $K$-user MARC with AF relaying, where all nodes except the receiver have multiple antennas. For this channel, we found the optimal solution for a previously introduced transmit strategy \cite{YH10}, where all transmitters access the channel jointly. Contrary to this approach, we proposed a transmit strategy that is based on a TDMA structure, where each transmitter accesses the channel exclusively. After optimizing the transmit covariance and relaying matrices for this scheme, the optimal time slot durations were found. Comparing the achievable sum-rates of the two strategies, it could be shown that the TDMA-based scheme always performs better in practical systems. For one specific scenario, a sum-rate gain of up to $10$\% was observed by simulations.

Although these results hold only for MARCs with a single-antenna receiver, it can be seen that the transmit strategy of \cite{YH10} is not generally optimal for the MARC. Hence, for the more general and unsolved MARC with a multi-antenna receiver, future research has to answer the question whether TDMA can help to increase the achievable sum rates.

%We have considered a $K$-user MARC, where all nodes except the receiver have multiple antennas. For this channel, we found the optimal solution for a previously introduced transmit strategy \cite{YH10}, where all transmitters access the channel jointly. Contrary to this approach, we also proposed a transmission strategy that is based on a TDMA structure, where each transmitter accesses the channel exclusively. As TDMA decomposes the $K$-user MARC in $K$ single-user channels, the transmit strategies of single-user relay channels can be used. After optimizing these strategies and finding the optimal time slot durations for the scheme, it could be shown that the achievable sum-rates of the TDMA-based scheme are larger or equal than the rates achievable by the previous scheme. Moreover, equality of the rates only occurs for cases that have probability zero in practical systems. Finally, a quantitative sum-rate gain of around $10$\% was observed for one specific scenario by simulations.

\section*{Acknowledgment}
The authors would like to thank Aydin Sezgin for fruitful initial discussions about the topic, which lead to the main ideas of this work. Moreover, the authors thank Karlheinz Ochs for providing a simplified proof of Lemma \ref{lem:pee_pi}.

%\appendices
\section*{Appendix A : Proof of Lemma \ref{lem:ev_pd}}
Let $\vmax$ be the eigenvector corresponding to the eigenvalue $\lmax\of{\B}$ with $\norm{\vmax} = 1$. Then, we have
\begin{align*}
\lmax\of{\B} &= \vmax^H\B\vmax\\
&\leq \vmax^H\B\vmax + \vmax^H\cdot(\A-\B)\cdot\vmax\\
&= \frac{\vmax^H\A\vmax}{\vmax^H\vmax}\\
&\leq \lmax(\A),
\end{align*}
where the first inequality is due to the positive semidefiniteness of $\A-\B$ and the last inequality can be derived from the Rayleigh quotient. \IEEEQED

\section*{Appendix B: Proof of Lemma \ref{lem:pee_pi}}
% % % FORMER PROOF % % %
%Defining $\B_0 = \A(P\I - \P)\A^H$, Lemma \ref{lem:pee_pi} can be proved by showing $\B_0 \succeq 0$. Using $\A=\U\Sg\V^H$ as the SVD of $\A$, we define $\Q = \V^H\P\V$. As $\Q$ is positive semidefinite and symmetric, we can diagonalize it as $\Q=\U_q\Sg_q\U_q^H$, where $\U_q$ is unitary and $\Sg_q = \diag(q_1,\ldots,q_{n})$ with $q_i\geq 0$ ($i=1,\ldots,n$). With these definitions, we will now derive matrices $\B_1$, $\B_2$, and $\B_3$ that are positive semidefinite if and only if $\B_0$ is positive semidefinite.

%First, let $\B_1 = \U^H\B_0\U = \Sg(P\I - \Q) \Sg^H.$ As $\U$ is unitary, $\B_0$ and $\B_1$ are similar matrices. Letting $\B_2 = P\I - \Q$ and using the fact that $\Sg$ is diagonal with only positive elements, it is easy to see that $\B_2 \succeq 0$ if and only if $\B_1 \succeq 0$. If $\B_3$ is defined as $\B_3 = \U_q^H\B_2\U_q = P\I - \Sg_q$, it is obviously similar to $\B_2$. As $\B_3$ is now a diagonal matrix, it is positive semidefinite if and only if its entries are non-negative, i.e., $P - q_i \geq 0$ ($i=1,\ldots,n$). As $\tr(\Sg_q) = \tr(\Q) = \tr(\P) \leq P$ and $q_i \geq 0$, this condition holds and the lemma is proved.

% % % NEW PROOF % % %
Lemma \ref{lem:pee_pi} can be equivalently formulated as
\begin{equation*}
\x^H\A(P\I - \P)\A^H\x \geq 0\;\;\; \forall \x \in \mathbb{R}^m.
\end{equation*}
Hence, it is sufficient to show that $P\I - \P$ is a positive semidefinite matrix. Let $\lambda_1,\ldots,\lambda_n$ be the eigenvalues of $\P$ with $\lambda_1 \geq \lambda_2 \geq \ldots \geq \lambda_n$. Then, we have
\begin{equation*}
P \geq \tr(\P) = \sum\limits_{i=1}^{n}\lambda_i \geq \lambda_1,
\end{equation*}
where the second inequality holds because we have $\P \succeq 0$ and thus $\lambda_i \geq 0\;\;\forall i$. Using $\lmin(P\I - \P)$ to denote the minimum eigenvalue of $P\I - \P$, we can write
\begin{equation*}
\lmin(P\I - \P) = P - \lmax(\P) = P - \lambda_1 \geq 0,
\end{equation*}
i.e., $P\I - \P$ is positive semidefinite.
\IEEEQED



%\bibliographystyle{IEEEtran}

\bibliographystyle{IEEEtran}


\end{document}